\documentclass[preprint]{imsart}
\RequirePackage[OT1]{fontenc}
\RequirePackage{amsthm,amsmath}
\RequirePackage[numbers]{natbib}
\RequirePackage[colorlinks,citecolor=blue,urlcolor=blue]{hyperref}
\usepackage{algorithmic, algorithm}
\usepackage{multirow} 
\usepackage{multicol} % Used for the two-column layout of the document
\usepackage{graphicx}
\usepackage{subfigure} 

\startlocaldefs
\numberwithin{equation}{section}
\theoremstyle{plain}
\newtheorem{thm}{Theorem}[section]
\endlocaldefs

\begin{document}

\begin{frontmatter}
\title{Diagnostics for Variational Bayes approximations}
\runtitle{Diagnostics for Variational Bayes approximations}

\author{\fnms{Hui} \snm{Zhao}\ead[label=e1]{h6zhao@uwaterloo.ca}}
\address{Department of Statistics and Actuarial Science,    University of Waterloo, 200 University Avenue West, Waterloo, Ontario, Canada N2L 3G1 \printead{e1}}

\and
\author{\fnms{Paul } \snm{Marriott}\ead[label=e2]{pmarriot@uwaterloo.ca}}
\address{Department of Statistics and Actuarial Science,    University of Waterloo, 200 University Avenue West, Waterloo, Ontario, Canada N2L 3G1 \\\printead{e2}}

\runauthor{Hui Zhao and Paul Marriott}

\begin{abstract}
 {\em Variational Bayes (VB)}   has shown itself to be a powerful approximation method in many application areas.  This paper describes some diagnostics methods which can  assess how well  the VB approximates the true posterior, particularly with regards to its covariance structure. The methods proposed  also allow us to  generate  simple corrections when the approximation error is large. It looks at  joint, marginal and conditional aspects of the approximate posterior and shows how to apply these techniques in both simulated and real data examples. 
\end{abstract}

\begin{keyword}
\kwd{Variational Bayes}
\kwd{Diagnostics}
\kwd{Covariance matrix}
\kwd{Independent Metropolis-Hastings}
\end{keyword}

\end{frontmatter}

\section{Introduction}

This paper looks at diagnostics tests to evaluate the quality of  {\em variational Bayes (VB)}  solutions. As  is typical with diagnostic testing in  statistics - think of diagnostic testing in regression analysis  for example - we  look at {\em necessary} conditions for adequacy.  A  VB solution may be inadequate from a number of perspectives. Here we list some of the  most important. Firstly,  by definition  the posterior covariance structure is distorted, so both posterior variances and correlations can be wrong.  Secondly,  VB convergence is only local, hence  it might miss other `better' solutions. In particular it  might focus on a single mode of a multimodal solution.  Thirdly,  there may be errors in higher order posterior moments, such as skewness or kurtosis.   A given diagnostic test is designed to detect a particular kind  of error, and the key idea of this paper  is to make available a number of computationally fast diagnostics, with a computation time of  the order of the VB itself,  which target particularly common forms of inadequacy,  specifically the first on the list above.

The VB method provides a fast and analytical approximation to otherwise intractable posterior distributions.  Early developments of the method can be found in  applications to  neural networks \cite{1993-Hinton&Camp}, \cite{1995-Mackay},  with further work in  independent component analysis \cite{Lappalainen99ensemblelearning}, \cite{Attias99independentfactor},  graphical models \cite{2003-Winn}, \cite{2003-Beal-Ghahramani},  information retrieval \cite{Blei03latentdirichlet}, and factor analysis \cite{2000-Ghahramani-Beal-FA}.    Other  applications of variational principle can be found in \cite{1991-Haff}, \cite{2009-McGrory-Titterington-Reeves-Pettitt}, \cite{2011-Faes-Ormerod-Wand}, \cite{2002-Hall-Humphreys-Titterington}, \cite{2006-Wang-Titterington},  while an excellent recent review   can be found in \cite{2010-Ormerod-Wand}. 

The essence of the method relies on making simplifying assumptions about the posterior dependence of a problem. This results in a   high dimension integral being  decomposed into a set of low dimensional ones which may be expected to be more tractable.  Various real-world applications have demonstrated that the VB method is very computationally efficient.  For example,   reversible jump MCMC \cite{Green95reversiblejump} may require many millions of  iterations to obtain posterior samples for a finite mixture of Normals, however,  the VB approximations can need only a few hundred, even though each iteration of the VB method is very fast.

For the examples of this paper, and others,  our studies show that VB can give good approximations to the posterior mean structure of a problem, and is good at finding  overall structural features -- such as the number of components in a mixture. However  the posterior variance can be underestimated.  This underestimation of the variance has also been reported by other researchers,  for example, \cite{2006-Bishop} and \cite{2009-Rue-Martino-Chopin}.  Moreover, by definition the general posterior dependence structure is distorted. This motivates the work in this paper to develop diagnostics to see  how well the VB approximations represent the actual posterior covariance structures, and to some extent to provide corrections when these errors are large. We emphasize  that these tests are only designed for these forms of error and may not detect errors of different kind. 

We  propose three diagnostics  methods which  only use the information obtained from VB approximations.   The first method   looks at the the joint  posterior distribution and attempts to find an optimal  affine transformation which links the VB and true posteriors.     The second method is based on a marginal posterior density approximation technique proposed by Tierney, Kass, Kadane (1989)\cite{1989-TKK}. Here  we work in specific low dimensional  directions to  estimate true posterior  variances and correlations.  The third  method, based on a stepwise conditional approach, allows us to   construct and solve a set of system of equations which lead to estimates of the true   posterior variances and correlations.   

This paper also proposes a novel method to calculate the variance of a marginal or conditional distributions of the posterior. This method uses an independent Metropolis-Hastings algorithm with the proposal kernel being configured by the VB approximations. Instead of using the sample moments,  the variance of the target distribution is computed by reading the acceptance probability of the generated MCMC chain. 

The paper is organized as follows. Section 2 presents the proposed methods in detail.  Applications of the methods  on the  simulated data and real-world data are shown in  Section 3.  Conclusions and discussions are  in Section 4.

\section{The three diagnostic methods}
 Consider the posterior distribution of a $p-$dimensional vector parameter $\theta =(\theta_{1},\cdots,\theta_{p})$, with  density function $p(\theta|x)$ where $x$ is an independent and identically distributed random sample.    We denote the VB approximation by $q(\theta)$,   the true posterior mean by $\mu = (\mu_{1},\cdots,\mu_{p})$,  the  covariance matrix by $\Sigma$ with variance of $\sigma^{2}_{i}, i=1,\cdots, p$ and correlation coefficients  $\{\rho_{ij}\}$.  
   
\subsection{Optimal affine transformations of joint distributions}

We denote the random vector associated with a VB approximation  by $\eta$.  We search for  optimal affine transformations  of $\eta$ (denoted by  $A \eta + B$, over specified classes of  $p\times p$ matrices, ($A$) and  $p \times 1$ vectors ($B$). The aim  is  to get  close to $\theta$,  the random vector associated with the true posterior.  

 First,  generate  an independent random sample of size $n$ from the VB distribution, denoted as $\{\eta_{i}\}_{i=1}^{n}$.  The values of $A$ and $B$ are obtained by maximizing a likelihood function, over the specified class, 
%{\setlength\arraycolsep{0.1em}
\begin{eqnarray}
Lik(A, B) := \prod_{i=1}^{n}(p( \theta_{i} |y; A, B) |\det(A)|), \label{eq:linearTrans}
\end{eqnarray} where $ \theta_{i} = A \eta_{i} + B,$  and $\det(A)$ is the determinant of $A$ and the corresponding estimates are  denoted by $\hat{A}$ and $\hat{B}$.
%}
Sampling from $q(\eta)$ is typically  straightforward since  $q(\eta)$ usually has a factorization form of $q(\eta) = \prod_{i}^{p} q_{i}(\eta_{i})$, and  $q_{i}(\eta_{i})$  often have  a well-known distributional form.   The maximization of (\ref{eq:linearTrans}) with respect to $A$ and $B$ is possible because it does not require the unknown normalizing constant of the posterior, $p(\theta|x)$.   

 For small or medium dimensional problems  we only   restrict  the transformation matrix $A$ to be   general lower triangular,  with the positive diagonal elements for identification reasons. For more complex problems sparser classes of matrices can be used, trading off the power of the test with speed.

\subsection{Marginal approximations}\label{sec:Marginal approximations}

This method considers a projection of the vector parameter $\theta$ in a direction $\alpha$, denoted by $\alpha^{T}\theta$.  The variance of  $\alpha^{T}\theta$ is given by $\alpha^{T}\Sigma\alpha$, which is a function of $\{\sigma_{i}^{2}\}_{i=1}^{p}$ and $\{\rho_{ij}\}$.  If we have the projections in different directions, we can obtain a system of equations which can be easily solved to obtain the values of $\{\sigma_{i}^{2}\}_{i=1}^{p}$ and $\{\rho_{ij}\}$.  

The key computation of this method is to calculate the value of the marginal variance.  In order to be  computationally efficient and exploit the VB solution we propose the following  new method. 

Suppose $p(\theta)$ is a target distribution and $q(\theta)$ is a proposal distribution.   The independent Metropolis-Hastings (IMH) algorithm will produce a transition from $\theta^{(t)}$ to $\theta^{(t+1)}$  as described in  Algorithm \ref{algm:IMH}.   Theorem \ref{thm: IMH-EAR-KL}, proved in {\cite{robert1999monte},  establishes a connection between  the {\em expected acceptance rate} (EAR) and the closeness of the target distribution $p(\theta)$ and the  proposal distribution $q$ 
measured in  Kullback-Leibler (KL) divergence. 
\begin{thm}
\label{thm: IMH-EAR-KL}
If there exists a constant M such that $p(\theta) < M q(\theta)$ for all $\theta$, then $\mbox{KL}(p||q) < \log(M)$ and the {\em expected acceptance rate (EAR)} is at least $\frac{1}{M}$ when the chain is stationary.
\end{thm}
Heuristically, Theorem \ref{thm: IMH-EAR-KL} states the closer  the target and the proposal, the higher the EAR. It is obvious that when $p(\theta)$ and $q(\theta)$ are identical, the optimal acceptance rate equals to one.  This result is different from  other types of Metropolis-Hastings  algorithms. For examples,  for random-walk Metropolis-Hastings algorithm the optimal acceptance rate is close to 0.234 \cite{1997-Roberts-Gelman-Gilks}; for Metropolis adjusted Langevin algorithms an overall acceptance rate is close to 0.574 \cite{1998-Roberts-Rosenthal-OptimalScaling}.  

Motivated by this general result,  first  consider  a special case in which the target distribution is a univariate normal with mean of $\mu$ and variance of $\sigma^{2}_{t}$ and the proposal distribution is a normal with the same mean and variance $\sigma^{2}_{p}$ (assume $\sigma^{2}_{p} > \sigma^{2}_{t}$).  It can be shown that  the EAR is monotone decreasing as the proposal variance of $\sigma^{2}_{p}$ increases.  Conversely, it says that given a fixed value of proposal variance of  $\sigma^{2}_{p}$, the value of the target variance $\sigma^{2}_{p}$ is one-to-one correspondence to the value of EAR.  This implies by monitoring the acceptance probability,  we can obtain the value of the target variance.   A table of expected acceptance rate versus the value of target variance  is given in the Appendix.

After this motivation let us consider the method in practice. Consider  two basics  facts:  firstly posteriors approach to normality when sample size is large, and secondly  VB provides good mean structure approximations. Hence we propose a new method to compute the target variance.  We call it a VB Adjusted Independent Metropolis-Hastings method (VBAIMH).    The variance of the target distribution is obtained by checking the acceptance rate for   a  standard normal kernel    centred at the VB mean,  being used   as the proposal.  In fact, the idea of using acceptance rates to compute the target variances can be further extended to using  acceptance rates as a key diagnostic to how close the VB distribution is to the true posterior. More discussion can be seen shortly.

The new approach  above has several advantages. First it does not require any particular tuning tricks to run the IMH algorithm.  We only need the posterior mean values produced by  the VB approximation to configure the proposal kernel.  This is a significant advantage over other MCMC methods,  in which  the implementation issues are the major concerns.

Secondly, this VB  kernel allows the MCMC chain to locate the regions of high posterior probability more efficiently, since the proposal kernel is around the posterior mode, at least locally;  then we only need a short chain to compute acceptance rates.  This is another significant advantage over the other MCMC methods, where the computational cost can be a big concern.  While the  IMH algorithm is well known  to perform poorly in high dimensions \cite{1998-geweke}, in this method  we are   sampling from a univariate distribution defined by the projections.  

Third, when an acceptance rate is low,  the generated sample may not represent the target distribution well, and the moments calculated upon these samples can be wrong. However, a low acceptance rate on it own reflects a significant discrepancy between the target and the proposal. 

\begin{algorithm}[!h] 
\caption{Independent  Metropolis-Hastings (IMH) algorithm}
\label{algm:IMH}
\begin{algorithmic}                    
\STATE Given $\theta^{(t)}$
\begin{description}
\item[Step 1] propose $\eta^{t} \sim q$.
\item[Step 2] Accept  $$\;\; \theta^{(t+1)} = \left\{ \begin{array}{ll}
         \eta^{t} & \mbox{with prob. $\min\left\{ \frac{p(\eta^{t})q(\theta^{(t)})}{p(\theta^{(t)})q(\eta^{t})}, 1  \right\}$ };\\
        \theta^{(t)} & \mbox{otherwise}.\end{array} \right. $$ 
\end{description}
\end{algorithmic}
\end{algorithm}

It is worth noting that when the projection is along the direction of a single parameter of $\theta_{i}$: that is, for example, we set $\alpha = (1, 0,\cdots, 0)$,  the VBAIMH can provide a possible mechanism to calibrate the EAR as a diagnostic tool to measure  inaccuracy in  the marginal approximations of VB, by using $q(\theta_{i})$ directly as the proposal distribution.   When the acceptance rate is low, it clearly indicates the approximation will be inaccurate.  Thus it gives us two uses: firstly a diagnostics tool in the general case and secondly it is possible to give  a correction to the VB approximation.  More discussion can be found in Section \ref{sec:Discussion}.

The above idea can be  further extended to  more general situations, where the diagnostics are targeted to  more specific errors.  For example, if only a subset of parameters is of immediate concern, which is  of particular usefulness in the high dimensional problems.  

To run an IMH algorithm we also need to know the density function of $\alpha^{T}\theta$, at least  up to scaling by  a normalizing constant.   Tierney, Kass, and Kadane (1989)\cite{1989-TKK} proposed an elegant marginal approximation of this  posterior distribution. 

Suppose the parameter of interest is  $\omega  = g(\theta)$,  where $g$ is a continuous real-valued function on $R^{p}$. The posterior distribution of $p(\omega|x)$ can be approximated as follows
 \begin{eqnarray}
p(\omega|x) \propto  \frac{\hat{p}(\omega|x)}{ | R_{\omega}|^{1/2} (b_{\omega}^{T}R_{\omega}^{-1}b_{\omega})^{1/2}}, 
 \end{eqnarray} where 
   \begin{eqnarray*}
\hat{p}(\omega|x) & =&  \operatorname*{sup}_{\theta: g(\theta)=\omega} p(\theta|x),  b_{\omega} = \frac{\partial g(\theta)}{\partial \theta}\Big|_{\theta=\theta_{\omega}},  R_{\omega} = \frac{\partial^{2} \log p(\theta|x)}{\partial \theta\theta^{T}}\Big|_{\theta=\theta_{\omega}}, \nonumber% \label{eq:TKK}
 \end{eqnarray*}
and  $\theta_{\omega}$ conditionally maximizes $p(\theta|x)$ with respect to $\theta$ for each given $\omega$. 

\subsection{The conditional stepwise method}

The unknown quantities in a true posterior covariance matrix $\Sigma$ are $\{\sigma_{i}^{2}\}_{i=1}^{p}$ and $\{\rho_{ij}\}$ and the difference between these and the VB versions which we are using as our diagnostics.  The stepwise method looks at conditional distributions constructed from the true posterior and compares them to ones based on the VB solution.   Algorithm \ref{algm:stepwise} gives a  description on the proposed method.    The three steps can be  explained as follows.  

Step 1 uses a linear transformation of $Y = Q \theta$ to scale the variances of $\{\sigma_{i}^{2}\}_{i=1}^{p}$ to be the ratios of $\{\sigma_{i}^{2}\}_{i=1}^{p}$ over their variational estimations, and uses a linear transformation of $Z = M Y$ to further scale these ratios to be one, which only leaves  $\{\rho_{ij}\}$ in $\Sigma$  to be  found. 

Step 2 finds a series of conditional bivariate random vector $U_{ij}$, to which the eigenvalues of their covariance matrix can be computed numerically after a rotation.  

Step 3 constructs a system of equations of $f_{k}(\{\rho_{ij}\})$ by linking the analytical expression of the correlation coefficient for the conditional bivariate $U_{ij}$, obtained based on the posterior normality (when sample size is large), to their numerical values of $r_{k}$, obtained by using the relationship between eigenvalues, and variances and correlation coefficients in a bivariate covariance matrix.  The values of $\{\sigma_{i}^{2}\}_{i=1}^{p}$ can be obtained by reversing Step 1. 

\begin{algorithm}[!h] 
\caption{The stepwise method }
\label{algm:stepwise}
\begin{algorithmic}                    
\STATE 
%\begin{description}
%\item[Step 1]
Step 1\\
\begin{itemize}
\item Define $Y = Q \theta$ and $\mu^{s} = Q\mu= (\mu^{s}_{1},\cdots,\mu^{s}_{p})$, where $Q = \left(\frac{1}{var_{q_{i}}(\theta_{i})}\right)$ is a diagonal matrix.  Denote $Y_{i}|Y_{-i}$ as the conditional $Y_{i}$  conditioning on  $Y_{j}= \mu^{s}_{j}, j\neq i$.  
\item Denote $m_{i}^{2} = var(Y_{i}|Y_{-i})$. Obtain $m_{i}^{2}, i=1,\cdots,p$, numerically. 
\item Define $Z = M Y$, and $\mu^{ss} = M\mu^{s}= (\mu^{ss}_{1},\cdots, \mu^{ss}_{p})$, where $M=\left(\frac{1}{m_{i}}\right)$ is a diagonal matrix.  .
\end{itemize}

%\item[Step 2]   
Step 2\\
\begin{itemize}
\item Denote $U_{ij} = Z_{ij}|Z_{-ij}, i \neq j$ as the conditional bivariate $(Z_{i}, Z_{j})$  conditioning on  $Z_{t}= \mu^{ss}_{t}, t \neq i,j$. 
\item Let $R =
\left(  \begin{array}{rr}
  \cos(\frac{\pi}{4}) & -\sin(\frac{\pi}{4}) \\
  \sin(\frac{\pi}{4})  & \cos(\frac{\pi}{4}) 
  \end{array}\right)$. Define $V_{k}$ = $(V_{k,1}, V_{k,2})$ = $RU_{ij}$,  where $k=1,\cdots, \frac{p(p-1)}{2}$ for all the pair of $i\neq j$. Denote $\lambda_{k,1}^{2} = var(V_{k,1})$ and $\lambda_{k,2}^{2} = var(V_{k,2})$. 
\item  Obtain  $\lambda_{k,1}^{2}$ and  $\lambda_{k,2}^{2}$, $k=1,\cdots,\frac{p(p-1)}{2}$, numerically.
  
\end{itemize} 

%\item[Step 3] 
Step 3\\
\begin{itemize}
\item Based on the posterior normality assumption,  compute the correlation coefficient for the conditional bivariate $U_{ij}$, and denote it as $f_{k}(\{\rho_{ij}\})$,  where $k=1,\cdots, \frac{p(p-1)}{2}$, for all the pair of $i\neq j$. 

\item  Compute $r_{k}  = \left(\frac{\lambda_{k,1}^{2}}{\lambda_{k,2}^{2}} -1\right) \big/ \left(\frac{\lambda_{k,1}^{2}}{\lambda_{k,2}^{2}} +1\right), k=1,\cdots,\frac{p(p-1)}{2}$. Solve the system of equations of $f_{k}(\{\rho_{ij}\}) = r_{k}$ to obtain the value of $\{\rho_{ij}\}$.

\item Based on the posterior normality assumption,  compute the conditional variance of $var(Y_{i}|Y_{-i})$, and denote it as $g_{i}(\sigma_{i}^{2})$, where $i=1,\cdots,p$.  Solve the equation $g_{i}(\sigma_{i}^{2}) = m_{i}^{2}$ to obtain the value of $\sigma_{i}^{2}$. 

\end{itemize} 
%\end{description}
\end{algorithmic}
\end{algorithm}

The key computations in  Algorithm \ref{algm:stepwise} involve computing the  values of a univariate conditional or marginal variances, that is $m_{i}^{2}, i=1,\cdots,p$ in Step 1 and $\lambda_{k,1}^{2}$ and  $\lambda_{k,2}^{2}$, $k=1,\cdots,\frac{p(p-1)}{2}$ in Step 2.   Again, these values can be computed by the VBAIMH method.   The definition of $r_{k}$ in Step 3 derives from the following fact.  For a bivariate  distribution, suppose the variances are $\sigma_{1}^{2}$ and $\sigma_{2}^{2}$ and correlation is $\rho$.  The  eigenvalues of covariance matrix are given as
$\lambda = \frac{(\sigma_{1}^{2}+\sigma_{2}^{2}) \pm \sqrt{(\sigma_{1}^{2}-\sigma_{2}^{2})^{2} + 4\rho^{2}\sigma_{1}^{2}\sigma_{2}^{2}  } }{2} $.  When $\sigma_{1}^{2}=\sigma_{2}^{2}$,  the eigenvalues are given by $
  \lambda_{1} = (1+\rho) \sigma^{2} $, and
$    \lambda_{2} = (1-\rho) \sigma^{2} 
$. Then it is easy to show that 
\begin{eqnarray}
\rho = \left(\frac{\lambda_{1}}{\lambda_{2}}-1\right)\big/ \left(\frac{\lambda_{1}}{\lambda_{2}}+1\right). \label{eq:eigen-rho}
\end{eqnarray} 

To illustrate  the method, we provide a three-dimension example which can be  found in the Appendix.

\section{Numerical studies}
In this section, we will work through four models with simulated or real datasets to  demonstrate the proposed methods.  For each model  we   compute  its variational approximation,  obtained by minimizing the the Kullback-Leibler (KL) divergence \cite{1951-Kullback-Leibler}. The distributional families of these approximations range widely; for example in the cases considered here they are  Normal, $t$, Beta,  Inverse Gamma, and Dirichlet.   

We start with a very basic  illustrative example: a large sample multivariate normal case with simulated data.  The second example looks at a  Normal random sample with unknown mean and variance with a real data set.  In this case  posterior normality is not assumed, showing that normality is not needed for the methods to have power.  We thirdly consider a two-component mixture of Normals model and finally a regime-switching lognormal model.  This last model can be considered a high-dimensional case with six interest parameters, and 528 latent nuisance parameters.  These models present a wide range of  complex dependence structures, and MCMC methods have been intensively studied with them; these models will provide good testimony for the proposed methods. For the regime-switching lognormal model, we used the real data set of the TSX monthly total return index in the period from January 1956 to December 1999, which contains 528 observations in total, see for discussion  \citep{2001-Hardy}, \citep{2002-Hardy}, and \cite{2011-Hartman-Heaton}. 

\subsection{Multivariate normal distributions}

We consider a 3-dimension vector parameter $\theta = (\theta_{1}, \theta_{2},\theta_{3})$,  and the posterior distribution of  $\theta$ and its VB approximation are all assumed to be a multivariate Normal distribution, and  these two normal distributions have the same mean values.  For the illustration purposes we will  arbitrarily choose the values of the covariance matrices for the true posterior and the  VB approximation. The following is an example; the actual variances are chosen to be $0.1^{2}$,  $1.3^{2}$,  and $4^{2}$ with correlation $0.51$ between $\theta_{1}$ and $\theta_{2}$, $0.37$ between $\theta_{1}$ and $\theta_{3}$, and $-0.3$.  In the VB approximation, the variances are assumed to be $\frac{0.1^{2}}{2.2}$, $\frac{1.3^{2}}{5.1}$, and $\frac{4^{2}}{6.9}$, and all correlation are assumed 0.

Our goal is that  given the covariance of VB approximations and the density function of the posterior distribution up to a normalizing constant we will compute the true covariance structure, more precisely,  to find the values of correlation coefficients of $0.51$, $0.37$, and $-0.3$ and the ratios of the posterior variances versus the VB variances, $2.2$, $5.1$, and $6.9$.

We first apply  the affine transformation method.  A sample of size of 600 is generated from the  VB distribution.  We restrict the transformation matrix $A$ to be a lower triangular matrix with positive diagonal elements.  There is no constraints on the three  parameters in the translation vector $B$.  Maximizing the posterior probability over these 9 parameters  can be done with the Newton's method or standard search methods.   The resulted $\hat{A}$ and  $\hat{B}$ are given by 
\begin{eqnarray}
\hat{A} = \left(  \begin{array}{ccc}
 1.527  & 0.000  & 0.000    \\
 10.498     &  2.007  &   0.000  \\
  23.601    &  -3.627 &   1.918
  \end{array}\right); 
 \hat{B} = \left(  \begin{array}{c}
  -0.004   \\
  -1.1089  \\
  1.758
  \end{array}\right)\nonumber \label{eq:}
\end{eqnarray} 
Given $\hat{A}$ and $\hat{B}$, the estimated $\hat{\Sigma}_{p}$ can be computed by $\hat{A}\Sigma_{v} \hat{A}^{T}$.    The second column in Table \ref{tab:mvn} gives the estimated variance ratios and correlation coefficients from $\hat{\Sigma}_{p}$.  We see these estimates are close to the actual values, which is given in  the first column.  
{\setlength\arraycolsep{0.1em}
\begin{table}[!h]
\centering
\caption{Multivariate Normal: 3 methods}
\label{tab:mvn}
\begin{tabular}{c| rrrr}
\hline
&True&Affine & Marginal & Stepwise\\
\hline
$\theta_{1}$ &$2.2$  &2.33    &2.54   &2.21     \\
$\theta_{2}$ &$5.1$  &5.54    &4.80   &5.04     \\
$\theta_{3}$ &$6.9$   &6.65   &7.04   &6.72         \\
$\rho_{12}$ & $0.51$ & 0.52  &0.56   &0.50     \\
$\rho_{13}$ &$0.37$  &  0.41 & 0.32   &0.40   \\
$\rho_{23}$ &$-0.30$&-0.24&-0.30    &-0.28\\
\end{tabular}
\end{table}

Second, we use  the method using  marginal approximations.  It requires 6 projections.  We denote a projection direction as $\alpha = (\alpha_{1}, \alpha_{2}, \alpha_{3})$, and the marginal variance along the projection direction as $l$.  Thus, we can obtain a polynomial equation involving $\sigma_{i}^{2}$ and $\rho_{ij}$  given by 
\begin{eqnarray*} &&\alpha_{1}^2 \sigma_{1}^{2} + \alpha_{2}^2 \sigma_{2}^{2} + \alpha_{3}^2 \sigma_{3}^{2}+ 2 \alpha_{1} \alpha_{2} \rho_{12} \sigma_{1}^{2} \sigma_{2}^{2} + 2 \alpha_{1} \alpha_{3} \rho_{13} \sigma_{1}^{2} \sigma_{3}^{2} + 2 \alpha_{2} \alpha_{3} \rho_{23} \sigma_{2}^{2} \sigma_{3}^{2}  = l\end{eqnarray*}
For each direction we simulate a sample of size 6000 and use the last 50\% sample points to calculate the acceptance rate. The values of $l$ (Table \ref{tab:mvn-mgnl}) are obtained from EAR table readings.   Solving the 6 polynomial equations, we obtained the following values; $\sigma_{1} = 1.59$,  $\sigma_{2} =  2.19 $, $\sigma_{3}  =  2.65   $,  $\rho_{12} = 0.56  $, $\rho_{13} = 0.32 $, $\rho_{23} = -0.30$.   The variance ratios are given in the third column in Table \ref{tab:mvn}.  We can see these estimates are consistent with those above and the true ones.

\begin{table}[!h]
\centering
\caption{The marginal approximation method: Multivariate Normal}
\label{tab:mvn-mgnl}
\begin{tabular}{lcc}
\hline
direction & Acceptance rate &  EAR reading: $l_{i}$ \\
$\frac{1}{3}(1,1,1)$ & 0.790 & 1.94\\
$\frac{1}{3}(1,-1,1)$ & 0.805 &  1.86\\
$\frac{1}{3}(1,1,-1)$ & 0.764&  2.12\\
$\frac{1}{3}(1,-1,-1)$ & 0.764&  0.47\\
$\frac{1}{3}(1,0.5,1)$ & 0.865 & 1.52 \\
$\frac{1}{3}(0.5,1.5,1)$ & 0.791&  1.94\\
\hline
\end{tabular}
\end{table}

Finally,  we work through the conditional stepwise method.  The notation used here follow that  given in Algorithm \ref{algm:stepwise}.  For Step 1 and 2, we simulate a sample of size 5000 and use the last 50\% sample points to calculate the acceptance rate, and the conditional variance is obtained from EAR  readings.   All numerical results are given in Table \ref{tab:mvn-stepwise}.  Solving the polynomial equations  obtains the values for $\rho_{12} = 0.50$, $\rho_{13} = 0.40$ and $\rho_{23} = -0.28$ and the  variance ratios, that are given in the fourth column in Table  \ref{tab:mvn}.  

{\setlength\arraycolsep{0.001em}
\begin{table}[h]
\caption{The stepwise method: Multivariate Normal}
\label{tab:mvn-stepwise}
\begin{tabular}{ p{0.1cm} | c  c c c  }
 &  Marginal variance   &  Acceptance rate  &  EAR table readings & Eigenvalue ratio \\ 
\hline %\\
\multirow{3}{*}{1}  
          & $m_{1}^{2}$ & 0.989  & 0.97 &  -\\ 
          & $m_{2}^{2}$ & 0.730  & 2.42 &  -\\ 
          & $m_{3}^{2}$ & 0.620  & 3.6  &  -\\ 
\hline %\\
\multirow{3}{*}{2}  
          & $\lambda_{1,1}^{2}, \lambda_{1,2}^{2}$ & 0.832, 0.651   & 0.588,  3.220 &  -\\ 
          & $\lambda_{2,1}^{2}, \lambda_{2,2}^{2}$ & 0.848, 0.681 &  0.602, 2.800 &-\\ 
          & $\lambda_{3,1}^{2}, \lambda_{3,2}^{2}$ & 0.718, 0.852  & 2.520, 0.625 &-\\ 
\hline %\\
\multirow{3}{*}{3}  
          & $r_{1}$ & -   & - &  0.691  \\ 
          & $r_{2}$ & -  &  - &  0.646  \\ 
          & $r_{3}$ & -  & -  &  -0.603  
\end{tabular}
\end{table}
}

\subsection{Normal random sample}

In this example, we consider a real dataset which contains 1033 records of weights for some Major League Baseball (MLB) Players \cite{2008-Onge}.  Plots suggests that it may be reasonable to model the data by a normal distribution with the  mean $\mu$ and variance $\sigma^{2}$.  We are interested in making inferences on $\mu$ and  $\sigma^{2}$.  In a Bayesian setting, we consider the priors as $\mu \sim N(\gamma, \eta^{2})$ and $\sigma^{2} \sim \mbox{IG}(\alpha, \beta)$, where $\mbox{IG}$ denotes the inverse Gamma distribution.  This setting is referred as a semi-conjugate prior \cite{1995-Gelman-et-al}.  The joint posterior distribution for $\mu$ and $\sigma^{2}$ is given by 
\begin{eqnarray}
p(\mu,\sigma^{2} |y) &\sim& \left(\frac{1}{\sigma^{2}}\right)^{-(\frac{n}{2}+\alpha+1)}  \exp\left(-\frac{1}{\sigma^{2}}\left(\frac{S^{2}}{2} + \beta + \frac{n(\mu-\bar{y})}{2}\right) - \frac{\mu-\gamma}{2\eta^{2}}\right),\label{eq:uvn-posterior-pdf}\nonumber
\end{eqnarray}   
where $\bar{y}$ is the sample mean and $S^{2}$ is the total sum of squares of $y$. The values of hyper-parameters are chosen to be $\alpha = 2$, $\beta = 440.64$,   $\gamma = 221.86$, and $\eta^{2} = 1$, where the values for $\beta$ and  $\gamma$ derives from the mean and variance of the dataset respectively.  

Despite the apparent  simplicity of this  model, the actual marginal posteriors for $\mu$ and $\sigma^{2}$ have no closed analytic forms.   We consider the VB  approximation, having a form of  $q(\mu,\sigma^{2}) = q(\mu)q(\sigma^{2})$.   VB  converges after 14 iterations.  The distributions for $q(\mu)$ and $q(\sigma^{2})$ are given as follows;
\begin{table}[htp]
%\begin{center}
\begin{tabular}{ p{2cm} |p{4.5cm} p{0.0001cm} }
\centering Parameter &  \centering Distribution (VB)  & \\ 
\hline %\\
\centering $\mu$ & \centering $N(208.09, 0.32)$  &\\
\centering $\sigma^{2}$ &  \centering$\mbox{IG}(518.50, 249154.70)$ &\\
\end{tabular}
\caption{The marginal distributions of VB approximations}
\label{tbl:UVN-marginal}
%\end{center}
\end{table}

\begin{table}[!h]
\caption{Posterior mean and covariance}
\label{tbl:uvn-post-mean-cov}
\begin{center}
\begin{tabular}{ c  |  c   | p{6.0cm}  p{0.0001cm}}
&  Posterior mean:  ($\mu$, $\sigma^{2}$) &  \centering Posterior covariance: ($\mu$, $\sigma^{2}$) & \\
\hline
\multirow{2}{*}{ Gibbs samples} & \multirow{2}{*}{ $( 208.10, 481.79)$}  &   \vtop{\vskip -7pt \vskip -\ht \strutbox  %\dp \ht
    \[ \left( \begin{matrix} 0.60^{2} & 0.34 \times0.60\times 22.64 \\
                                      0.34 \times0.60\times 22.64     &  22.64^{2} \end{matrix} \right)\] \vskip -5pt\vskip  -\ht\strutbox } & \\
\hline
\multirow{2}{*}{ VB approx.} & \multirow{2}{*}{ $( 208.09, 481.46)$}  &   \vtop{\vskip -0pt \vskip -\ht \strutbox  %\dp \ht
    \[ \left( \begin{matrix} 0.56^{2} &   0 \\
                                      0     &  21.18^{2} \end{matrix} \right)\] \vskip -8pt\vskip  -\ht\strutbox } & \\
\hline
{ Ratios } & ( 1, 1)  &  \centering(1.13, 1.14) &
\end{tabular}
\end{center}
\end{table}

As a standard comparison, we run  a Gibbs sampler, and simulate a sample of size $10^{5}$ from the posterior distribution.  Table \ref{tbl:uvn-post-mean-cov} gives a comparison of posterior mean and posterior covariance  estimated by  the VB approximation and the MCMC sample moments.  The third row in Table \ref{tbl:uvn-post-mean-cov} gives the ratios of posterior mean and the ratios of posterior variance.  We can see that the means estimated by both methods  are almost  identical.  However the variances  approximated by VB are slightly underestimated, (as expected from our discussion above)  and it is obvious VB distorts the posterior dependence structure.   

We applied the three proposed methods to this  problem. All the setting and routines used here are very similar to the those used in the previous example.  The numerical results produced by each method are given in the Appendix.  The final results for three methods are given in Table  \ref{tab:uvn-3methods-estimates}.  We see that  all of the  methods perform well.  

\begin{table}[!h]
\caption{MLB Players weights: 3 methods}
\label{tab:uvn-3methods-estimates}
\centering
\begin{tabular}{l | cc  p{0.1cm} c}
\hline
& \multicolumn{2}{c}{Variance ratios} & & \multicolumn{1}{c}{Correlation coeff. } \\
& $\mu$ & $\sigma^{2}$   && $\rho$  \\
\cline{1-1} \cline{2-3}  \cline{5-5} 
Gibbs &  $1.13$ & $1.14$     &&  $0.34$\\
Affine  & 1.09   & 1.13   &&  $0.30$\\
Marginal  & 1.08 & 1.20  &&  $0.35$ \\
Stepwise  &  1.13 &1.14   &&  $0.35$%\\
%\hline
\end{tabular}
\end{table}

\subsection{Finite mixture models}\label{sec:Finite mixture models}

In this example, we  considers a two-component mixture of normals.  The density function is given by
$f(x_{i}|\Psi) = \pi  \phi(x_{i}|\mu_{1}, \sigma^{2}_{1} ) + (1-\pi)  \phi(x_{i}|\mu_{2}, \sigma^{2}_{2} ), \label{eq:ch1-normal-mixture}$  
where $0< \pi<1$,  $\phi \text{ is the normal density function}$, $\Psi = (\pi, \mu_{1}, \sigma^{2}_{1},\mu_{2}, \sigma^{2}_{2})$ are the model parameters.    Based on the conjugacy consideration, We choose the following prior distributions, 
\begin{eqnarray*}
p(\pi) &= &\mbox{Beta}\left(\frac{a_{0}}{2},\frac{a_{0}}{2}\right), a_{0} > 0; \; \\
p(\mu_{j}, \sigma^{2}_{j}) &=  & \displaystyle \mathrm{N}\bigg( \mu_{j}| \sigma_{j}^{2};c_{j},\frac{\sigma_{j}^{2}}{d_{j}^{2}}\bigg)\mathrm{IG}( \sigma_{j}^{2};e_{j}, f_{j}), j = 1,2.\nonumber
\end{eqnarray*}
where $a_{0}$, $c_{j}$, $d_{j}^{2}$, $e_{j}$, $f_{j}$ are hyper-parameters.   Given a dataset $x=\{x_{i}\}_{i=1}^{n}$, the posterior distribution is given by 
$$p(\pi, \mu_{1}, \sigma^{2}_{1},\mu_{2}, \sigma^{2}_{2}|x) \propto \prod_{i=1}^{n} f(x_{i}|\Psi) p(\pi)p(\mu_{1}, \sigma^{2}_{1})p(\mu_{2}, \sigma^{2}_{2}).$$
We consider such a VB approximation having a factorization as  
$$q(\pi, \mu_{1}, \sigma^{2}_{1},\mu_{2}, \sigma^{2}_{2}) = q(\pi) q(\mu_{1}|\sigma^{2}_{1}) q(\sigma^{2}_{1})q(\mu_{2}|\sigma^{2}_{2}) q(\sigma^{2}_{2}).$$

\begin{figure}[!h]
  \caption{The histogram of the dataset for a two component mixtures of Normals}
  \label{fig:mixture-data}
  \centering
  \includegraphics[width=0.4\textwidth]{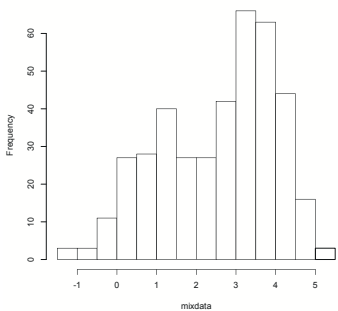}
\end{figure}
For illustration,  we consider a special model  $$f(x) = 0.4  \phi(x;1,1)+ 0.6 \phi(x;3.5,0.5).$$  A sample dataset of size 400 was  generated from this model.  Figure \ref{fig:mixture-data} shows the histogram of the dataset.   The VB method for this dataset converges after 122 iterations.   The approximation distributions are given in Table \ref{tbl:mixture-VB}. 

\begin{table}[!htp]
%\begin{center}
\begin{tabular}{ p{2cm} |p{4.5cm} p{0.0001cm} }
\centering Parameter &  \centering Distribution (VB)  & \\ 
\hline %\\
\centering $\pi$ & \centering $\mbox{Beta}(167.35, 232.67)$  &\\
\centering $\mu_{1}|\sigma^{2}_{1}$ & \centering $\mbox{N}(1.13,  \sigma^{2}_{1}/168.34)$  &\\
\centering $\sigma_{1}^{2}$ &  \centering$\mbox{IG}( 85.66, 76.56)$ &\\
\centering $\mu_{2}|\sigma^{2}_{2}$ & \centering $\mbox{N}(3.57,  \sigma^{2}_{2}/233.66)$  &\\
\centering $\sigma_{2}^{2}$ & \centering $\mbox{IG}(118.33, 50.31)$  &\\ 
\end{tabular}
\caption{The marginal distributions of  VB approximations}
\label{tbl:mixture-VB}
%\end{center}
\end{table}
Alternatively, we run  a Gibbs sampler, and simulate a sample of size $2\times10^{6}$ from the posterior distribution.  The posterior mean and covariance  estimated by  the VB approximations and by the MCMC sample moments  are given in Table \ref{tbl:mx-post-mean} and \ref{tbl:mx-post-cov}.  The ratios of posterior means and the  ratios of posterior variances are given in the  last row in Table \ref{tbl:mx-post-mean} and \ref{tbl:mx-post-cov}.  These ratios indicate  that the means estimated by both methods  are almost  identical.  However VB underestimated the actual posterior variances,  and again strongly distorts the correlation structure. 
\begin{table}[!h]
\caption{Posterior means: 2-component mixtures of Normals }
\label{tbl:mx-post-mean}
\begin{tabular}{ c |r  }
&  Post. mean:  ($\pi$, $\mu_{1}$, $\mu_{2}$, $\sigma^{2}_{1}$, $\sigma^{2}_{1}$)\\
\hline
Gibbs samples & $(0.418, 1.126, 3.566, 0.885, 0.430)$ \\
VB approx. &  $(0.418, 1.131, 3.571, 0.904, 0.429)$ \\
\hline
{ Ratios } & $(1.000, 0.996, 0.999, 0.979, 1.003) $
\end{tabular}
\end{table}
\begin{table}[!h]
\caption{Posterior covariance: 2-component mixtures of Normals }
\label{tbl:mx-post-cov}
\begin{tabular}{ c   | p{8cm}  }
& Post. covariance: ($\pi$, $\mu_{1}$, $\mu_{2}$, $\sigma^{2}_{1}$, $\sigma^{2}_{1}$)\\
\hline
\multirow{6}{1.2cm}{ Gibbs samples} &   variance:  \\
&  (0.00208, 0.01902, 0.00349, 0.03696, 0.00431) \\
\cline{2-2}
&  correlation coeff.:  \\
& (0.83,  0.81,  0.94,  0.69,  0.84, 0.73,  $-0.60$,  $-0.66$,  $-0.76$, $-0.50$) \\
\hline
\multirow{6}{1.2cm}{ VB approx.}  & variance:  \\
& (0.00061, 0.00537, 0.00183, 0.00977, 0.00158)  \\
\cline{2-2}
&  correlation coeff.: \\
& (0.00, 0.00, 0.00, 0.00, 0.00, 0.00, 0.00, 0.00, 0.00, 0.00 )  \\
\hline
{ Ratios } &  (3.42, 3.54, 1.90, 3.78, 2.73)
\end{tabular}
\end{table}

The proposed methods are applied to this mixture problem.  For the VBAIMH method, we generate 4000 samples to compute acceptance rates.  The sample size is much smaller than is used in the Gibbs sampler, to which we use a large sample size to ensure the chain has in fact converged.

In addition, we note that in this example we have targeted the most general form of linear transformation to correct any inadequacy in posterior mean and variance. In fact a diagnostic test can be designed to be targeted at more special concerns. For example the matrix $A$ in the transformation might be restricted to be of a particular class such as  diagonal or banded.  This would be particular useful in high dimensional problems where the dimension of the space of $A$  could become problematic.

All numerical results associated with each method are given in the Appendix.  The final results are given in  Table \ref{tab:mx-3methods-estimates}.  We see all the estimates are close to the values computed by using MCMC samples (the first column).  

\begin{table}[!h]
\caption{Two-component mixtures of Normals: 3 methods}
\label{tab:mx-3methods-estimates}
\begin{tabular}{c|rrrr}
\hline
&Gibbs &Affine  & Marginal  & Stepwise  \\
$\pi$  &  3.42   & 3.64   &   5.56   &    3.06   \\
$\mu_{1}$ & 3.54    & 4.51  &   4.73   & 4.12   \\
$\mu_{2}$&    1.90  &   3.53  &  2.37  & 2.70   \\
$\sigma^{2}_{1}$  &3.78  & 3.35 & 4.15  & 3.21\\
$\sigma^{2}_{2}$ & 2.73 & 2.62 & 2.76 &  2.56     \\
$\rho_{12}$ &   0.83   &  0.76   &   0.63   & 0.71   \\
$\rho_{13}$ &   0.81   &  0.68   &    0.59  & 0.63   \\
$\rho_{23}$ &   0.94   &  0.71   &   0.87   & 0.65   \\
$\rho_{14}$  &   0.69  &  0.66   &   0.47   & 0.64   \\
$\rho_{24}$  &   0.84  &  0.73   &    0.82  & 0.70    \\
$\rho_{34}$  &   0.73  &  0.59   &  0.76    & 0.53   \\
$\rho_{15}$  &  -0.60  &  -0.61  & -0.46    &  -0.58    \\
$\rho_{25}$  &  -0.66  &  -0.61  & -0.66   & -0.58   \\  
$\rho_{35}$   & -0.76  & -0.63   & -0.71   & -0.60   \\
$\rho_{45}$  &  -0.50 & -0.50    &  -0.54  & -0.47
\end{tabular}
\end{table}

\subsection{The regime-switching model}\label{sec:The regime-switching model}

Our final example  considers  the high-dimensional  regime-switching lognormal model (RSLN),  proposed by \cite{2001-Hardy}, which is used to  model the switching processes  between different states or regimes in  many time series.   For this example we   consider  a real dataset of the TSX monthly total return index in the period from January 1956 to December 1999 (528 observations in total).   This dataset is studied  in \citep{2001-Hardy}, \citep{2002-Hardy}, and \cite{2011-Hartman-Heaton} in the context of maximum likelihood and MCMC methods.  We offer an VB solution for this model in \citep{2013-zhao-rsln}.   

The regime-switching lognormal  model   with a fixed finite number, K, of regimes can be described as a bivariate discrete time process with the observed data sequence $w_{1:T} = \{w_{t}\}_{t=1}^{T}$ and the unobserved regime  sequence $S_{1:T} = \{S_{t}\}_{t=1}^{T}$, where $S_{t} \in \{1,\cdots,K\}$ and $T$ is the number of observations. This unobserved   sequence forms the high dimensional latent structure for this problem. The logarithm of $w_{t}$, denoted by $y_{t} = \log w_{t}$, is assumed normally distributed having   mean $\mu_{i}$ and variance $\sigma_{i}^{2}$ both dependent on the hidden regime $S_{t}$.   The  sequence of $S_{1:T}$  is assumed to follow a first order Markov chain having transition probabilities $A = (a_{ij})$ with the probabilities $\pi = (\pi_{i})_{i=1}^{K}$ to  start the first regime.  

In the Bayesian framework,  we use a symmetric Dirichlet prior for $\pi$, that is 
$$p(\pi) = \mbox{Dir}(\pi; \frac{C^{\pi}}{K},\cdots,\frac{C^{\pi}}{K}), \text{ for }C^{\pi} > 0.$$  
Let $a_{i}$ denote the $i^{th}$ row vector of A.  The prior for $A$ is chosen as 
$$p(A) =  \prod_{i=1}^{K} p(a_{i})  =  \prod_{i=1}^{K} \mbox{Dir}(a_{i}; \frac{C^{A}}{K},\cdots,\frac{C^{A}}{K}),\text{ for } C^{A} >0.$$ 
The prior distribution for $\{(\mu_{i},\sigma^{2}_{i})\}_{i=1}^{K}$ is chosen normal-inverse gamma, 
$$p(\{\mu_{i},\sigma^{2}_{i}\}_{i=1}^{K})=\prod_{i=1}^{K} \mathrm{N}( \mu_{i}| \sigma_{i}^{2};\gamma, \frac{\sigma_{i}^{2}}{\eta^{2}})\mathrm{IG}( \sigma_{i}^{2};\alpha, \beta).$$  
In the above setting,  $C^{\pi}$, $C^{A}$, $\gamma$, $\eta^{2}$, $\alpha$, and $\beta$ are hyper-parameters.  Thus,  the joint posterior distribution of $\pi, A, \{\mu_{i},\sigma^{2}_{i}\}_{i=1}^{K},  \text{ and } S_{1:T}$ can be obtained as, 
{\setlength\arraycolsep{0.001em}
\begin{eqnarray}
P(\pi, A, \{\mu_{i},\sigma^{2}_{i}\}_{i=1}^{K},S_{1:T}|y_{1:T}) \propto && p(S_{1}|\pi)\prod_{t=1}^{T-1}p(S_{t+1}| S_{t};A)\prod_{t=1}^{T}p(y_{t}|S_{t}; \{\mu_{i},\sigma^{2}_{i}\}_{i=1}^{K}) \nonumber\\
&&p(\pi)p(A)p(\{\mu_{i},\sigma^{2}_{i}\}_{i=1}^{K})  \label{eq:joint-post}.
\end{eqnarray}
}
we  choose the variational approximation which is factorized as follows
{\setlength\arraycolsep{0.001em}
\begin{eqnarray*}
&&q(\pi, A, \{\mu_{i},\sigma^{2}_{i}\}_{i=1}^{K},S_{1:T}) =  q(\pi)\prod_{i=1}^{K}q( a_{i})\prod_{i=1}^{K}q(\mu_{i}|\sigma^{2}_{i})q(\sigma^{2}_{i})q(S_{1:T})\label{eq:joint-vb}.
\end{eqnarray*}
}
In \citep{2013-zhao-rsln}, we show that VB suggests a two-regime RSLN model for the monthly TSX total return data. The VB approximations are given as follows:  
\begin{table}[!htp]
%\begin{center}
\begin{tabular}{ p{2cm} |p{4.5cm} p{0.0001cm} }
\centering Parameter &  \centering Distribution (VB)  & \\ 
\hline %\\
\centering $\mu_{1}$ & \centering $t_{454.61}(0.0123, 370778.19)$  &\\
\centering $\sigma_{1}^{2}$ &  \centering$\text{IG}(227.30, 0.28)$ &\\
\centering $\mu_{2}$ & \centering $t_{80.39}( -0.0161,  12987.55)$  &\\
\centering $\sigma_{2}^{2}$ & \centering $\text{IG}(40.20, 0.24)$  &\\ 
\centering $a_{1,2}$ & \centering $\text{Dirichlet}(15.21, 434.78)$  &\\ 
\centering $a_{2,1}$ & \centering $\text{Dirichlet}(15.00,  61.21)$  &
\end{tabular}
\caption{The marginal distributions of  VB approximations}
\label{tbl:tsx-marginal}
%\end{center}
\end{table}

The posterior mean and covariance  estimated by  the VB approximations and by the MCMC sample moments (cited from \cite{2002-Hardy})  are given in Table \ref{tbl:rsln-post-mean} and \ref{tbl:rsln-post-cov}.  The ratios of posterior means is given in the  last column in Table \ref{tbl:rsln-post-mean},  and the  ratios of posterior variances in the last row in Table  \ref{tbl:rsln-post-cov}.  These ratios indicate  that the means estimated by both methods  are almost  identical.  However VB underestimated the actual posterior variances,  and again strongly distorts the correlation structure.

\begin{table}[!h]
\caption{Posterior means: the regime-switching lognormal  model}
\label{tbl:rsln-post-mean}
\begin{tabular}{ c |ccc }
Post. mean:  & Gibbs & VB    & Ratios \\
\hline
$\mu_{1}$       & 0.0122 &  0.0123  & 0.991\\
$\sigma_{1}$  & 0.0351 &  0.0349  & 1.006   \\
$a_{1,2}$        & 0.0334 &  0.0338  & 0.988 \\
$\mu_{2}$       & -0.0164 & -0.0161 & 1.0198 \\
$\sigma_{2}$  & 0.0804 &  0.0777  &  1.035\\
$a_{2,1}$        & 0.2058  & 0.1969  & 1.045
\end{tabular}
\end{table}
\begin{table}[!h]
\caption{Posterior covariance: the regime-switching lognormal  model}
\label{tbl:rsln-post-cov}
\begin{tabular}{ c  | p{9cm}  }
& Post. cov.: ($\mu_{1}$,  $\sigma_{1}$,$a_{1,2}$, $\mu_{2}$, $\sigma_{2}$,$a_{2,1}$)\\
\hline
\multirow{6}{1.2cm}{ Gibbs samples} &   s.d.:  \\
&  0.002, 0.002, 0.012, 0.010,0.009, 0.065 \\
\cline{2-2}
&  correlation coeff.:  \\
& -0.16, 0.17, -0.34,-0.10, -0.11, 0.08, -0.17, 0.22,-0.25,-0.15,
0.06,-0.04,0.34,-0.14,0.12 \\
\hline
\multirow{6}{1.2cm}{ VB approx.}  & s.d.:  \\
& 0.0017, 0.00008, 0.0085, 0.0089, 0.0010, 0.045  \\
\cline{2-2}
&  correlation coeff.: \\
& 0, 0, 0, 0, 0, 0, 0, 0, 0, 0,0, 0, 0, 0, 0  \\
\hline
{ Ratios } &  1.22, 24.46,1.47,  1.19, 8.21,  1.46
\end{tabular}
\end{table}

The proposed methods are applied to this RSLN model.  Again,  we use 4000 samples to compute the acceptance rates in VBAIMH, which significant shorten the  computational time, compared with other MCMC methods.  

In this example, we work through a complete cycle of the stepwise method to obtain the exact estimations on variances and correlation coefficients, which provide a quantitative correction on VB approximations.  In fact, each step of the stepwise method can provide a qualitative diagnostics. Table \ref{tab:rsln-stepwise} gives all numerical results in the stepwise method.  The values of $m_{i}$ in Step 1 are all greater than 1, which indicates that the VB variances are smaller than the true ones, since a conditional variance always penalizes the marginal variance.  In Step 2,   the values of the pair of $\lambda_{k,1}$ and $\lambda_{k,2}$ provide possible information about the sign of the correlation.  The final results are given in  Table \ref{tab:rsln-3methods-estimates}.

\begin{table}[!h]
\caption{The regime-switching lognormal  model: 3 methods}
\label{tab:rsln-3methods-estimates}
\begin{tabular}{c|rrrr}
\hline
&Gibbs &Affine  & Marginal  & Stepwise  \\
$\mu_{1}$ 		      & 1.22    & 1.44    &  1.97    & 1.22    \\
$\sigma^{2}_{1}$    & 24.46  & 1.60    &  1.76    &    1.63\\
$a_{1,2}$                &1.47    &  2.81   &   1.41 &     2.21 \\
$\mu_{2}$              &  1.19    & 1.20   &    1.82    &   1.33  \\
$\sigma^{2}_{2}$    & 8.21   &  1.57    &   2.00  &      1.49      \\
$a_{2,2}$                &1.46    &  2.34   &   2.15  &     1.96\\
\hline
$\rho_{12}$ & -0.1630    &   -0.1217     &    -0.1175    &-0.1266     \\
$\rho_{13}$ & 0.1681     &     0.2228    &     0.1220 &  0.1367   \\
$\rho_{23}$ & -0.3438    &   -0.2970     &    -0.3831 &  -0.3388   \\
$\rho_{14}$  & -0.1043   &   -0.1294     &    -0.1874 &  -0.1275  \\
$\rho_{24}$  & -0.1094   &    -0.0903     &   -0.0649 &  -0.0865     \\
$\rho_{34}$  & 0.0796    &    0.0221     &    0.0856  & 0.0507   \\
$\rho_{15}$  &  -0.1678  &   -0.1856     &    -0.1061  &  -0.1328   \\
$\rho_{25}$  &  0.2235   &    0.1793     &   0.1008  &  0.1390    \\  
$\rho_{35}$   & -0.2517  &   -0.1604     &   -0.2890  & -0.2160    \\
$\rho_{45}$  &  -0.1476  &   -0.0747      &  -0.0116 &  -0.0231 \\
$\rho_{16}$  &  0.0552   &    0.0528     &   0.0942  &  0.0640   \\
$\rho_{26}$  &  -0.0374   &  -0.0690     &   0.0461  &  -0.0772  \\  
$\rho_{36}$   & 0.3385   &   0.3985      &    0.5947  &  0.3518  \\
$\rho_{46}$  &  -0.1433  &  -0.1154      &   -0.1664  &  -0.0989  \\
$\rho_{56}$  & 0.1238    &   0.1291      &   0.1434  &  0.1023
\end{tabular}
\end{table}

\section{Discussion}\label{sec:Discussion}

The variational method essentially provides posterior marginal approximations,  which can be inaccurate in a number of ways.  The present paper aims to provide fast and easy-to-use diagnostics, which mainly target on inadequacy in the covariance structure.    From the above numerical studies we can see all three  methods can provide both  diagnostics showing the quality of the VB approximation and also,  in these examples, good estimates on the actual posterior variances and correlations. These methods are easy to use.  They are free of any  sophisticated tuning techniques or special expertise and   are  highly  computational efficient compared with the traditional sampling based methods.   

This paper introduces a novel way to use acceptance rates.  The idea is that the acceptance rate can act as a key diagnostic to how close the VB distribution is to the true posterior.   As discussed in Section \ref{sec:Marginal approximations}, EAR could be calibrated as a diagnostics tool to measure the inadequacy in marginal approximations by using VB approximations directly as the proposal distributions  in VBAIMH.  For the situation when  posteriors depart from normality,  a low acceptance rate still indicate an inaccurate approximation.  However, to quantify a particular form of  inaccuracy, a single value of EAR  may be diluted by the confounding of many factors: inadequate variance, inadequate skewness, inadequate tail behaviour.  In the further research, separating these confounding factors would be a necessary step toward measuring a special form of  inadequacy.

For high dimensional problems, the three proposed methods can be designed to target  more specific situations. For example, the covariance matrix might be  sparse; a subset of the parameters might be of  immediate concern.  As discussed in Section \ref{sec:Marginal approximations}, Section \ref{sec:Finite mixture models}, and Section \ref{sec:The regime-switching model},   the three methods can offer different strategies to address the special form of diagnostics.  In the affine method, the transformation matrix $A$  might be restricted to be a particular class.  In the marginal method,  the projections might be set to particular directions.  In the stepwise method,  the  steps might be applied to a subset of the parameters, conditioning on other parameters.

For each individual method, we have some further comments.  The affine transformation based method relies on using  approximate  linear relationship between the VB approximation and the actual posterior as a diagnostics and potentially a correction.   In cases where,  for example,  strong skewness is  present in the posterior the correction will of course not be exact, but it will still be a useful diagnostics tool. 

As  Leonard, Hsu and Tsui (1989)\cite{1989-Leonard} point out, the marginal approximation of Tierney, Kass, and Kadane (1989), primarily justified by asymptotically $n\rightarrow \infty$, might be insufficient for  finite $n$; they also show a number of examples in which the method of Tierney, Kass, and Kadane  introduces excessive skewness in the marginal approximations.  We also find the inadequacy in the method of Tierney, Kass, and Kadane in our numerical studies. For example:  Table \ref{tab:mx-mgnl-variance} gives the values of 15 marginal variances along 15 directions in the mixture problem. The first row is the analytical results calculated from  the covariance matrix obtained from a Gibbs samples.  The second row is based on the posterior marginal approximation. We can see some discrepancy between this two sets of numbers.  Even though,  the method based on the marginal approximation still works well.  Leonard, Hsu and Tsui (1989)\cite{1989-Leonard} proposed a refinement on the method of Tierney, Kass, and Kadane (1989), that could be considered in  future work.
\begin{table}[!h]
\caption{The marginal variances by marginal approximations: mixtures of Normal}
\label{tab:mx-mgnl-variance}
{\setlength\arraycolsep{0.001em}
\begin{tabular}{p{0.7cm}|p{0.5cm}p{0.5cm}p{0.5cm}p{0.5cm}p{0.5cm}p{0.5cm}p{0.5cm}p{0.5cm}}
\hline
 & $l_{1}$ & $l_{2}$ & $l_{3}$& $l_{4}$& $l_{5}$& $l_{6}$& $l_{7}$& $l_{8}$\\
Gibbs  & 3.415  & 0.590   & 1.312   & 0.801   & 6.630   &  0.916  & 0.672   & 0.659 \\
Aprx. & 3.703 & 0.999  & 1.854  & 1.31  & 4.818  & 1.394  & 1.052  & 0.912 \\
\hline
 & $l_{9}$& $l_{10}$& $l_{11}$& $l_{12}$& $l_{13}$& $l_{14}$& $l_{15}$\\
Gibbs   & 0.495 & 2.510  &  0.650  & 0.712  &  1.976   & 0.436 & 2.987  \\
Aprx.   & 0.792  & 3.376  & 0.915  & 1.133  & 2.477  & 0.634  & 2.96 \\
%\hline
\end{tabular}
}
\end{table}

Another potential concern when using  the marginal approximation method is that  we need to perform a constrained  maximization at each sampling step.  The maximization can often be completed in straightforward fashion such as the standard Newton's method or search methods. However a optimization at each sampling step may affect the computational efficiency.  

The VBAIMH provides a fast means to calculate the variance of the target distribution.  When using  this approach two particular issues arise.  First,  as discussed previously when the target variance of $\sigma^{2}_{t}$ is greater than the proposal variance of $\sigma^{2}_{p}$, the EAR is monotone decreasing as $\sigma^{2}_{t}$ increases.  Similarly,   when $\sigma^{2}_{t} < \sigma^{2}_{p}$ the EAR is also monotone decreasing as $\sigma^{2}_{t}$ decreases.   This means that we need to determine if  $\sigma^{2}_{t} < \sigma^{2}_{p}$ or not,  before determining the value of  $\sigma^{2}_{t}$.  In practice,  we can assume  $\sigma^{2}_{t} > \sigma^{2}_{p}$ and pick a value of $\sigma^{2}_{t}$ from the EAR table, and then use this new value as the proposal variance and run the IMH again.  If the new acceptance rate is close to one or increases,  this means $\sigma^{2}_{t} > \sigma^{2}_{p}$ otherwise $\sigma^{2}_{t} < \sigma^{2}_{p}$.   If $\sigma^{2}_{t} < \sigma^{2}_{p}$ is the case,  the true value of $\sigma^{2}_{t}$ is the reciprocal of the value read from the EAR table.

Second, if the approximate normality of posterior distributions does not hold well, for example in the cases where strong  skewness is present,  the variance read from the EAR table will confound these non-normality effects and will deviate from the true value.  As discussed above, when we pick a value of $\sigma^{2}_{t}$ from the EAR table  and  use this new value as the proposal variance to run another IMH algorithm,  if the new acceptance rate is not close to one, this implies that the normality does not hold well and usually it is skewness presents.  When this happens,  we need to adjust the value read from the EAR table.  We usually scale the readings as $c \sigma^{2}_{t}$.   Based on our various numerical studies,  a reasonable choice on the scale $c$ is 0.85.

\section{Appendix} 

\subsection{Stepwise method: a 3-dimension example}

All the notations used here are defined in  Algorithm \ref{algm:stepwise}.  We consider a vector parameter  $\theta = (\theta_{1},\theta_{2},\theta_{3})$, and denote its posterior variance and correlation as $\sigma_{1}^{2}$, $\sigma_{2}^{2}$  and  $\sigma_{3}^{2}$, and  $\rho_{1}$, $\rho_{2}$ and $\rho_{3}$

{ Step 1.} Define $s_{i}^{2} = \sigma_{i}^{2}/var_{q_{i}}(\theta_{i}), i=1,2,3$.  Then, by the linear transformation of $Y=Q\theta$, the variances of $Y$ are given by $ s_{1}^{2}$, $ s_{2}^{2}$, and $ s_{3}^{2}$ respectively, with the correlation coefficients of $\rho_{1}$, $\rho_{2}$ and $\rho_{3}$ unchanged.  Based on the posterior normality conditions, the conditional variance of $var(Y_{1}| Y_{2}, Y_{3})$, $var(Y_{2}| Y_{1}, Y_{3})$, and $var(Y_{3}| Y_{2}, Y_{1})$ are given respectively 
{\setlength\arraycolsep{0.1em}
 \begin{eqnarray}
var(Y_{1}| Y_{2}, Y_{3})&=& (1- \frac{\rho_{1}^{2}+\rho_{2}^{2}- 2\rho_{1}\rho_{2}\rho_{3} }{1-\rho_{3}^{2}})s_{1}^{2} = m_{1}^{2}   \label{eq:m-1} \\
var(Y_{2}| Y_{1}, Y_{3}) &=& (1- \frac{\rho_{1}^{2}+\rho_{3}^{2}- 2\rho_{1}\rho_{2}\rho_{3} }{1-\rho_{2}^{2}})s_{2}^{2} = m_{2}^{2}  \label{eq:m-2}\\
var(Y_{3}| Y_{2}, Y_{1}) &=& (1- \frac{\rho_{2}^{2}+\rho_{3}^{2}- 2\rho_{1}\rho_{2}\rho_{3} }{1-\rho_{1}^{2}})s_{3}^{2} = m_{3}^{2}  \label{eq:m-3}
 \end{eqnarray}
}
The value of $m_{1}^{2}$,  $m_{2}^{2}$, and $m_{3}^{2}$ can be obtained numerically by using the VBAIMH algorithm.  After linear transformation of $Z = M Y$, the variance of $Z_{1}$, $Z_{2}$, and $Z_{3}$ are given respectively by
{\setlength\arraycolsep{0.1em}
 \begin{eqnarray*}
var(Z_{1})&=&  \frac{1-\rho_{3}^{2}}{1-(\rho_{1}^{2}+\rho_{2}^{2}+\rho_{3}^{2})- 2\rho_{1}\rho_{2}\rho_{3}}, \\
var(Z_{2})&=&  \frac{1-\rho_{2}^{2}}{1-(\rho_{1}^{2}+\rho_{2}^{2}+\rho_{3}^{2})- 2\rho_{1}\rho_{2}\rho_{3}}, \\
var(Z_{3})&=& \frac{1-\rho_{1}^{2}}{1-(\rho_{1}^{2}+\rho_{2}^{2}+\rho_{3}^{2})- 2\rho_{1}\rho_{2}\rho_{3}}, 
\end{eqnarray*}
}where   only $\rho_{1}^{2}$, $\rho_{2}^{2}$, and $\rho_{3}^{2}$ are involved.

{ Step 2.}  There are three bivariate random vectors in total in $Z$:  $U_{12} = (Z_{1}, Z_{2}|Z_{3})$, $U_{13} = (Z_{1}, Z_{3}|Z_{2})$, and $U_{23} = (Z_{2}, Z_{3}|Z_{1})$.   The two random variables in  $U_{12}$ have equal variances, similar for $U_{13}$, and $U_{23}$.  By the eigen-decomposition, the covariance matrix of $U_{12}$ can be expressed as 
$var(U_{12}) = R^{T} \Bigg(  \begin{array}{cc}
  \lambda_{1} & 0 \\
  0  & \lambda_{2} 
  \end{array}\Bigg) R$, where R is the rotation matrix defined in Algorithm \ref{algm:stepwise}, and $\lambda_{1}$ and $\lambda_{2}$ are the eigenvalues of $var(U_{12})$. Thus,  the covariance matrix of $V_{1}$ is diagonal with $\lambda_{1}$ and $\lambda_{2}$ as the entries. The values of $\lambda_{1}$ and $\lambda_{2}$ can be computed numerically, by running the VBAIMH algorithm.

{Step 3.}  Based on the posterior normality conditions,  the correlation coefficient $r_{1}$ of  $U_{12}$, $r_{2}$ of $U_{13}$, and $r_{3}$ of $U_{23}$ are given respectively by  
{\setlength\arraycolsep{0.1em}
 \begin{eqnarray}
r_{1} &=& \frac{(\rho_{1}-\rho_{2}\rho_{3})}{\sqrt{(1-\rho_{3}^{2})(1-\rho_{1}^{2})}}  \label{eq:3dmath-step3-1}  \\
r_{2} &=& \frac{(\rho_{2}-\rho_{1}\rho_{3})}{\sqrt{(1-\rho_{3}^{2})(1-\rho_{1}^{2})}}  \label{eq:3dmath-step3-2}\\
r_{3} &=& \frac{(\rho_{3}-\rho_{1}\rho_{2})}{\sqrt{(1-\rho_{1}^{2})(1-\rho_{1}^{2})}} \label{eq:3dmath-step3-3}
\end{eqnarray}
}

Given the values of $\lambda_{1}$ and $\lambda_{2}$ in Step 2, we can obtain  the value of $r_{1}$ by computing $r_{1}  = \left(\frac{\lambda_{1}^{2}}{\lambda_{2}^{2}} -1\right) \big/ \left(\frac{\lambda_{1}^{2}}{\lambda_{2}^{2}} +1\right)$; similar to compute $r_{2}$ and $r_{3}$.   Thus,  we can obtain and solve a system of three polynomial equations given in (\ref{eq:3dmath-step3-1}), (\ref{eq:3dmath-step3-2}), and (\ref{eq:3dmath-step3-3}) to obtain the values of $\rho_{1}$,$\rho_{2}$, and $\rho_{3}$; further, the value of $s_{1}^{2}$, $s_{2}^{2}$, and $s_{3}^{2}$ can be obtained by solving (\ref{eq:m-1}), (\ref{eq:m-2}), and (\ref{eq:m-3}), and then the value of $\sigma_{1}^{2}$, $\sigma_{2}^{2}$, and $\sigma_{3}^{2}$.

\subsection{Numerical results for the example of Normal random sample}
For the affine transformation method, the resulted $\hat{A}$ and  $\hat{B}$ is given by,
\begin{eqnarray}
\hat{A} = \left(  \begin{array}{cc}
 1.049  & 0.000 \\
 11.824     &  1.018 
  \end{array}\right); 
 \hat{B} = \left(  \begin{array}{c}
  -10.119    \\
  -2469.079 
  \end{array}\right).\nonumber \label{eq:}
\end{eqnarray} 

For  the marginal approximation method, the directional vectors with the corresponding acceptance rates and EAR table readings are given in  Table \ref{tab:uvn-mgnl}.   
\begin{table}[!h]
\centering
\caption{The Method upon on marginal approximations: MLB Players weights}
\label{tab:uvn-mgnl}
\begin{tabular}{lcc}
\hline
direction & Acceptance rate &   EAR reading: $l_{i}$ \\
$\frac{1}{\sqrt{2}}(1,1)$ & 0.861 & 1.55\\
$\frac{1}{\sqrt{2}}(1,-1)$ & 0.901 &  0.74\\
$\frac{1}{\sqrt{2}}(1,0.5)$ & 0.966&  0.893\\
\hline
%\hline
\end{tabular}
\end{table}

The numerical results for  the stepwise method for each step are given in Table \ref{tab:uvn-stepwise}.   
\begin{table}[h]
\caption{The stepwise method: MLB Players weights}
\label{tab:uvn-stepwise}
\begin{center}
\begin{tabular}{ c | c  c c c  }
%\vspace{0.05in}	
 &  Variance   &  Acceptance rate  &  EAR readings & Ratio \\ 
\hline %\\
\multirow{2}{*}{1}  
          & $m_{1}^{2}$ & 0.991   & 1.03  &  -\\ 
          & $m_{2}^{2}$ & 0.966  & 1.13 &  -\\ 
\hline %\\
\multirow{1}{*}{2}  
          & $l_{1,1}^{2}, l_{1,2}^{2}$ & 0.881,  0.889   & 0.694, 1.44  &  -\\ 
\hline %\\
\multirow{1}{*}{3}  
          & $r_{1}$ & -   & - &  2.075 \\ 
\end{tabular}
\end{center}
\end{table}

\subsection{Numerical results for mixture of Normals model}
For the affine transformation method,  the estimated $\hat{A}$ and  $\hat{B}$ is given by,

{\setlength\arraycolsep{0.1em}
\begin{eqnarray}
\hat{A} = \left(  \begin{array}{ccccc}
 1.907  & 0.000  &  0.000   &  0.000  &  0.000 \\
  4.822   & 1.372   & 0.000  & 0.000  & 0.000 \\ 
  2.227   & 0.329   & 1.255  & 0.000  & 0.000 \\ 
  4.867  &  0.862  &  0.235  &  1.210  & 0.000 \\ 
  -1.593 &  -0.190 &  -0.342 & -0.011  & 1.177 
  \end{array}\right); 
 \hat{B} = \left(  \begin{array}{c}
  -0.368   \\
 -2.399 \\
 -2.199 \\
 -4.003  \\
  2.026  
  \end{array}\right).\nonumber \label{eq:}
\end{eqnarray} 
}
For  the marginal approximation method,  it requires 15 projections.  The directional vectors with the corresponding acceptance rates and EAR table readings are given in  Table \ref{tab:mx-mgnl}.
\begin{table}[!h]
\centering
\caption{The marginal approximation method: mixtures of Normals}
\label{tab:mx-mgnl}
\begin{tabular}{lcc}
\hline
direction & Acceptance rate &  EAR reading: $l_{i}$ \\
$\frac{1}{3}(1, 1, 1, 1, 1)$ & 0.582 & 4.120\\
$\frac{1}{3}(1, -1, 1, 1, 1)$ & 0.901 & 0.735 \\
$\frac{1}{3}(1, 1, -1, 1, 1)$ & 0.815  & 1.800 \\
$\frac{1}{3}(1, 1, 1, -1, 1)$ &  0.906 & 1.340\\
$\frac{1}{3}(1, 1, 1, 1, -1)$ & 0.477  &  6.500\\
$\frac{1}{3}(-1, 1, 1, 1, 1)$ & 0.882  &  1.420\\
$\frac{1}{3}(-1,-1, 1, 1, 1)$ & 0.923  &  1.280 \\
$\frac{1}{3}(-1, 1, -1, 1, 1)$ &0.944   & 0.840\\
$\frac{1}{3}(-1, 1, 1, -1, 1)$ & 0.876 &  0.676 \\
$\frac{1}{3}(-1, 1, 1, 1, -1)$ &  0.619 & 3.600 \\
$\frac{1}{3}(1, -1, -1, 1, 1)$ & 0.940 & 0.830 \\
$\frac{1}{3}(1,-1, 1, -1, 1)$ &  0.944 & 1.200 \\
$\frac{1}{3}(1, -1, 1, 1, -1)$ &  0.727 & 2.380 \\
$\frac{1}{3}(1, 1, -1, -1, 1)$ &  0.830 & 0.581\\
$\frac{1}{3}(1, 1, -1, 1, -1)$ &  0.633 &  3.400\\
\hline
%\hline
\end{tabular}
\end{table}

For the stepwise method.  The numerical results for Step 1 and 2  are given in Table \ref{tab:mx-stepwise}.  
\begin{table}[h]
\caption{The stepwise method: mixtures of Normals}
\label{tab:mx-stepwise}
\begin{center}
\begin{tabular}{ p{0.1cm}  | c  c c c  }
%\vspace{0.05in}	
 &  Variance   &  Acceptance rate  &  EAR readings & Ratio \\ 
\hline %\\
\multirow{5}{*}{1}  
          & $m_{1}^{2}$ & 0.933   & 1.230 &  -\\ 
          & $m_{2}^{2}$ & 0.885  &  1.428 &  -\\ 
          & $m_{3}^{2}$ & 0.923  & 1.286 &  -\\ 
          & $m_{4}^{2}$ & 0.817  & 1.512 &  -\\ 
          & $m_{5}^{2}$ & 0.853  & 1.425 &  -\\ 
\hline %\\
\multirow{10}{*}{2}  
          & $\lambda_{1,1}^{2}, \lambda_{1,2}^{2}$ &  0.917, 0.879  & 0.776, 1.462  &  -\\ 
          & $\lambda_{2,1}^{2}, \lambda_{2,2}^{2}$ &  0.935, 0.919  & 0.816, 1.282  &  -\\ 
          & $\lambda_{3,1}^{2}, \lambda_{3,2}^{2}$ &  0.888, 0.848  & 0.803, 1.303  &  -\\ 
          & $\lambda_{4,1}^{2}, \lambda_{4,2}^{2}$ &  0.885, 0.906  & 1.221, 0.827  &  -\\ 
          & $\lambda_{5,1}^{2}, \lambda_{5,2}^{2}$ &  0.930, 0.911  & 0.802, 1.331 &  -\\ 
          & $\lambda_{6,1}^{2}, \lambda_{6,2}^{2}$ &  0.857, 0.795  & 0.701, 1.660  &  -\\ 
          & $\lambda_{7,1}^{2}, \lambda_{7,2}^{2}$ &  0.885, 0.921  & 1.174, 0.858  &  -\\ 
          & $\lambda_{8,1}^{2}, \lambda_{8,2}^{2}$ &  0.906, 0.880  & 0.887, 1.090  &  -\\ 
          & $\lambda_{9,1}^{2}, \lambda_{9,2}^{2}$ &  0.857, 0.890  & 1.356, 0.754  &  -\\ 
          & $\lambda_{10,1}^{2}, \lambda_{10,2}^{2}$ & 0.978, 0.822   & 0.984, 0.934  &  -\\ 
\hline %\\
\multirow{10}{*}{3}  
          & $r_{1}$ & -   & - &  0.307   \\ 
          & $r_{2}$ & -   & - &  0.222   \\ 
          & $r_{3}$ & -   & - &  0.238 \\ 
          & $r_{4}$ & -   & - &  -0.192  \\ 
          & $r_{5}$ & -   & - &  0.248   \\ 
          & $r_{6}$ & -   & - &  0.406  \\ 
          & $r_{7}$ & -   & - &  -0.155  \\ 
          & $r_{8}$ & -   & - &   0.103 \\ 
          & $r_{9}$ & -   & - &   -0.285  \\ 
          & $r_{10}$ & -   & - &  -0.026 \\ 
\end{tabular}
\end{center}
\end{table}

\newpage
\subsection{Numerical results for the regime-switching lognormal model}
For the affine transformation method,  the estimated $\hat{A}$ and  $\hat{B}$ is given by,

{\setlength\arraycolsep{0.1em}
\begin{eqnarray*}
\hat{A} = \left(  \begin{array}{rrrrrr}
 1.200  &   &     &    &   \\
  -0.008   & 1.257   &   &   &  \\ 
   1.864   &  -45.563  & 1.570 &   &  \\ 
  -0.722  &   -11.953  &  0.026 & 1.076  &  \\ 
  -0.156   &   2.657  &  -0.342 & -0.014   & -0.013  &1.208 \\ 
   2.195   &  -52.512   &   3.292  &   -1.028  &   11.013  &   1.361\\
  \end{array}\right); 
 \hat{B} = \left(  \begin{array}{r}
  -0.002    \\
 0.000  \\
 0.017   \\
 0.024  \\
  -0.002 \\
  -0.220  
  \end{array}\right).
  \end{eqnarray*} 
}

For  the marginal approximation method,  it requires 21 projections.  The directional vectors with the corresponding acceptance rates and EAR table readings are given in  Table \ref{tab:rsln-mgnl}.
\begin{table}[!h]
\centering
\caption{The marginal approximation method: the regime-switching lognormal model}
\label{tab:rsln-mgnl}
\begin{tabular}{lcc}
\hline
direction & Acceptance rate &  EAR reading: $l_{i}$ \\
$\frac{1}{3}(1, 1, 1, 1, 1,1)$ & 0.852   & 1.6   \\
$\frac{1}{3}(1,-1, 1, 1, 1,1)$ & 0.811   &  1.84  \\
$\frac{1}{3}(1, 1,-1, 1, 1, 1)$ & 0.843   & 1.66 \\
$\frac{1}{3}(1, 1, 1,-1, 1,1)$ & 0.867   &  1.50 \\
$\frac{1}{3}(1, 1, 1, 1,-1,1)$ & 0.755   &  2.16 \\
$\frac{1}{3}(1, 1, 1, 1, 1,-1)$ & 0.845   &  0.62 \\
$\frac{1}{3}(-1, 1, 1, 1, 1, 1)$ & 0.873  &  1.48 \\
$\frac{1}{3}(-1,-1, 1, 1, 1, 1)$ &0.897   &  1.36 \\
$\frac{1}{3}(-1, 1,-1, 1, 1, 1)$ & 0.877   &    1.48 \\
$\frac{1}{3}(-1, 1, 1,-1, 1, 1)$ & 0.884  &  1.44   \\
$\frac{1}{3}(-1, 1, 1, 1,-1, 1)$ & 0.873  &  1.48    \\
$\frac{1}{3}(-1, 1, 1, 1, 1, -1)$ &  0.849 & 0.64   \\
$\frac{1}{3}(1,-1,-1, 1, 1, 1)$ &  0.817   & 0.56   \\
$\frac{1}{3}(1,-1, 1,-1, 1, 1)$ & 0.841   & 1.60   \\
$\frac{1}{3}(1,-1, 1, 1,-1, 1)$ & 0.819   & 1.76   \\
$\frac{1}{3}(1,-1, 1, 1, 1, -1)$ &  0.890  &0.73  \\
$\frac{1}{3}(1, 1,-1,-1, 1, 1)$ & 0.802    & 1.52  \\
$\frac{1}{3}(1, 1,-1, 1,-1, 1)$ &  0.865   & 0.65 \\
$\frac{1}{3}(1, 1,-1, 1, 1, -1)$ &  0.854  & 1.62   \\
$\frac{1}{3}(1, 1, 1,-1,-1, 1)$ &  0.768   & 2.08 \\
$\frac{1}{3}(1, 1, 1,-1, 1, -1)$ &   0.865 & 0.65\\
\hline
%\hline
\end{tabular}
\end{table}

For the stepwise method.  The numerical results for Step 1 and 2  are given in Table \ref{tab:rsln-stepwise}.  
\begin{table}[h]
\caption{The stepwise method: the regime-switching lognormal model}
\label{tab:rsln-stepwise}
\begin{center}
\begin{tabular}{ p{0.1cm}  | c  c c c  }
%\vspace{0.05in}	
 &  Variance   &  Acceptance rate  &  EAR readings & Ratio \\ 
\hline %\\
\multirow{5}{*}{1}  
          & $m_{1}^{2}$ &  0.951 & 1.15 &  -\\ 
          & $m_{2}^{2}$ &  0.862 & 1.42 &  -\\ 
          & $m_{3}^{2}$ &  0.817  & 1.61 &  -\\ 
          & $m_{4}^{2}$ &  0.928  & 1.28 &  -\\ 
          & $m_{5}^{2}$ &  0.807  & 1.35 &  -\\ 
          & $m_{6}^{2}$ &  0.831 & 1.63 &  - \\
\hline %\\
\multirow{10}{*}{2}  
          & $\lambda_{1,1}^{2}, \lambda_{1,2}^{2}$ & 0.929, 0.936   & 1.29, 0.91  &  -\\ 
          & $\lambda_{2,1}^{2}, \lambda_{2,2}^{2}$ & 0.949, 0.931   &  0.97, 1.30 &  -\\ 
          & $\lambda_{3,1}^{2}, \lambda_{3,2}^{2}$ & 0.950, 0.960   & 1.18, 0.89 &  -\\ 
          & $\lambda_{4,1}^{2}, \lambda_{4,2}^{2}$ & 0.844, 0.875   & 1.61, 0.90  &  -\\ 
          & $\lambda_{5,1}^{2}, \lambda_{5,2}^{2}$ & 0.966, 0.962   & 0.98, 1.14 &  -\\ 
          & $\lambda_{6,1}^{2}, \lambda_{6,2}^{2}$ & 0.871, 0.873   & 1.51, 0.77  &  -\\ 
          & $\lambda_{7,1}^{2}, \lambda_{7,2}^{2}$ & 0.921, 0.943   & 1.27, 0.91 &  -\\ 
          & $\lambda_{8,1}^{2}, \lambda_{8,2}^{2}$ & 0.921, 0.774   & 0.88, 2.09  &  -\\ 
          & $\lambda_{9,1}^{2}, \lambda_{9,2}^{2}$ & 0.952, 0.907   & 0.99, 1.04  &  -\\ 
          & $\lambda_{10,1}^{2}, \lambda_{10,2}^{2}$ & 0.926, 0.930 & 0.94, 1.29  &  -\\ 
          & $\lambda_{10,1}^{2}, \lambda_{11,2}^{2}$ & 0.858, 0.819 & 1.53, 0.77  &  -\\ 
          & $\lambda_{11,1}^{2}, \lambda_{12,2}^{2}$ & 0.917, 0.783 & 0.76, 2.00  &  -\\ 
          & $\lambda_{13,1}^{2}, \lambda_{13,2}^{2}$ & 0.866, 0.884   & 0.96, 0.96  &  -\\ 
          & $\lambda_{14,1}^{2}, \lambda_{14,2}^{2}$ & 0.930, 0.961   & 1.19, 0.94  &  -\\ 
          & $\lambda_{15,1}^{2}, \lambda_{15,2}^{2}$ & 0.889, 0.810   & 0.82, 1.22 &  -\\ 
 \hline %\\
\multirow{10}{*}{3}  
          & $r_{1}$ & -   & - & -0.127   \\ 
          & $r_{2}$ & -   & - &  0.137     \\ 
          & $r_{3}$ & -   & - & -0.338    \\ 
          & $r_{4}$ & -   & - &  -0.128  \\ 
          & $r_{5}$ & -   & - & -0.087      \\ 
          & $r_{6}$ & -   & - & 0.057    \\ 
          & $r_{7}$ & -   & - & -0.133    \\ 
          & $r_{8}$ & -   & - &  0.139  \\ 
          & $r_{9}$ & -   & - &  -0.216   \\ 
          & $r_{10}$ & -   & - & -0.023   \\ 
          & $r_{11}$ & -   & - & 0.064     \\ 
          & $r_{12}$ & -   & - & -0.077     \\ 
          & $r_{13}$ & -   & - & 0.352  \\ 
          & $r_{14}$ & -   & - & -0.099     \\ 
          & $r_{15}$ & -   & - & 0.102    \\ 
\end{tabular}
\end{center}
\end{table}

\subsection{The EAR table}

The EAR table, shown in Table  \ref{tab:EAR table 1},  is composed as follows:  the label for rows contains the first two digits of the target variance; the label for columns contains the decimal of the target variance;  the values within the table are expected acceptance rates.  For example:  if one obtains an acceptance rate of 0.5555, then one would look for the rows to find 4 and the columns to 0.6 which yields the target variance is 4.6.

\begin{center}
\begin{table*}[h!]
\caption{EAR table:  variance versus expected acceptance rate}
\label{tab:EAR table 1}
\begin{tabular}{  c | c  c c  c  c c  c  c c  c  }
\hline\hline	
  &0 & 0.1 & 0.2 & 0.3 & 0.4 & 0.5 & 0.6 & 0.7 & 0.8 & 0.9\\ 
\hline
1 & 1.0000 & 0.9697 & 0.9422 & 0.9165 & 0.8936 & 0.8720 & 0.8517 & 0.8330 & 0.8157 & 0.7990\\
 2 & 0.7833 & 0.7690 & 0.7553 & 0.7423 & 0.7299 & 0.7182 & 0.7067 & 0.6960 & 0.6860 & 0.6761\\
3 & 0.6671 & 0.6577 & 0.6492 & 0.6406 & 0.6325 & 0.6249 & 0.6175 & 0.6104 & 0.6034 & 0.5969\\
4 & 0.5903 & 0.5842 & 0.5782 & 0.5719 & 0.5664 & 0.5609 & 0.5555 & 0.5502 & 0.5450 & 0.5405\\
5 & 0.5354 & 0.5306 & 0.5261 & 0.5221 & 0.5174 & 0.5131 & 0.5089 & 0.5049 & 0.5011 & 0.4971\\ \vspace{-0.35cm}\\

6 & 0.4937 & 0.4897 & 0.4861 & 0.4827 & 0.4791 & 0.4758 & 0.4725 & 0.4693 & 0.4660 & 0.4634\\
7 & 0.4602 & 0.4571 & 0.4542 & 0.4513 & 0.4485 & 0.4460 & 0.4431 & 0.4404 & 0.4376 & 0.4355\\
8 & 0.4325 & 0.4304 & 0.4279 & 0.4251 & 0.4232 & 0.4207 & 0.4184 & 0.4162 & 0.4139 & 0.4118\\
9 & 0.4098 & 0.4077 & 0.4056 & 0.4034 & 0.4013 & 0.3998 & 0.3975 & 0.3956 & 0.3936 & 0.3916\\
10 & 0.3900 & 0.3883 & 0.3864 & 0.3849 & 0.3829 & 0.3809 & 0.3794 & 0.3776 & 0.3759 & 0.3744\\\vspace{-0.35cm}\\

11 & 0.3728 & 0.3713 & 0.3698 & 0.3682 & 0.3664 & 0.3651 & 0.3638 & 0.3620 & 0.3608 & 0.3592\\
12 & 0.3581 & 0.3563 & 0.3550 & 0.3538 & 0.3520 & 0.3509 & 0.3497 & 0.3486 & 0.3474 & 0.3458\\
13 & 0.3444 & 0.3432 & 0.3419 & 0.3408 & 0.3396 & 0.3384 & 0.3370 & 0.3361 & 0.3346 & 0.3337\\
14 & 0.3325 & 0.3311 & 0.3302 & 0.3292 & 0.3277 & 0.3271 & 0.3257 & 0.3250 & 0.3235 & 0.3226\\
15 & 0.3217 & 0.3209 & 0.3196 & 0.3185 & 0.3177 & 0.3165 & 0.3157 & 0.3148 & 0.3137 & 0.3127\\ \vspace{-0.35cm}\\

16 & 0.3121 & 0.3108 & 0.3100 & 0.3090 & 0.3082 & 0.3074 & 0.3064 & 0.3054 & 0.3047 & 0.3039\\
17 & 0.3030 & 0.3023 & 0.3013 & 0.3005 & 0.2993 & 0.2989 & 0.2979 & 0.2970 & 0.2964 & 0.2954\\
18 & 0.2947 & 0.2940 & 0.2930 & 0.2924 & 0.2915 & 0.2909 & 0.2901 & 0.2894 & 0.2886 & 0.2877\\
19 & 0.2870 & 0.2861 & 0.2854 & 0.2852 & 0.2844 & 0.2834 & 0.2827 & 0.2822 & 0.2814&0.2808\\
\hline
\end{tabular}
\end{table*}
\end{center}

\pagebreak

\bibliographystyle{imsart-nameyear}
\bibliography{stats-diagnostics.bib}

\end{document}